\documentclass[11pt]{article}
\usepackage{amsmath,amssymb,color,graphics,epsfig,cite}

\usepackage{amsfonts}

\newcommand{\be}{\begin{equation}}
\newcommand{\ee}{\end{equation}}
\newcommand{\bea}{\setlength\arraycolsep{2pt} \begin{eqnarray}}
\newcommand{\eea}{\end{eqnarray}}
\newcommand{\nn}{\nonumber}

\def\ft#1#2{{\textstyle{\frac{\scriptstyle #1}{\scriptstyle #2} } }}
\def\fft#1#2{{\frac{#1}{#2}}}

\def\0{{\sst{(0)}}}
\def\1{{\sst{(1)}}}
\def\2{{\sst{(2)}}}
\def\3{{\sst{(3)}}}
\def\4{{\sst{(4)}}}
\def\5{{\sst{(5)}}}
\def\6{{\sst{(6)}}}
\def\7{{\sst{(7)}}}
\def\8{{\sst{(8)}}}
\def\sst#1{{\scriptscriptstyle #1}}

\usepackage{xcolor}

\def\tr{\rm tr}
\usepackage{multirow}

\begin{document}

\begin{center}
{\Large {\bf Tightening the Penrose Inequality}}

\vspace{20pt}
{\large H.~Khodabakhshi, H.~L\"u and Run-Qiu Yang*}

\vspace{10pt}

{\it Center for Joint Quantum Studies and Department of Physics\\ School of Science, Tianjin University,\\ Yaguan Road 135, Jinnan District, Tianjin 300350, China}
\vspace{40pt}

\underline{ABSTRACT}
\end{center}

The Penrose inequality estimates the lower bound of the mass of a black hole in terms of the area of its horizon. This bound is relatively loose for extremal or near extremal black holes. We propose a new Penrose-like inequality for static black holes involving the mass, area of the black hole event horizon and temperature. Our inequality includes the Penrose inequality as its corollary, and it is saturated by both the Schwarzschild and Reissner-Nordstr\"om black holes. In the spherically-symmetric case, we prove this new inequality by assuming both the null and trace energy conditions.
\\
~\\
~\\
Keywords: Black hole, ~~Penrose-like inequality,  ~~trace energy condition\\
PACS: 04.70.Bw, ~~04.30.-w, ~~04.20.-q

\vfill{\footnotesize  h.khodabakhshi@ipm.ir \ \ \ mrhonglu@gmail.com
\ \ \ aqiu@tju.edu.cn}

\thispagestyle{empty}
\pagebreak



\section{Introduction}
Some physical principles state that spacetime will approach an asymptotically stationary black hole state  at late times \cite{Wald:1999vt}. Given the initial Cauchy data with a horizon $\sigma$, if the cosmic censorship conjecture is valid \cite{Penrose:1969pc,Hawking:1970zqf,4}, i.e. the future singularity is surrounded by the event horizon and the spacetime is in a stationary state, then the black hole no-hair or uniqueness theorem \cite{5} will indicate that the final state should be the Kerr metric \cite{Chrusciel:2008js}. The surface area of the Kerr horizon is
\be\label{kerr}
A_h=8\pi M_{\rm B} \big(M_{\rm B}+\sqrt{M_{\rm B}^2-a^2}\big)\le 16\pi M_{\rm B}^2\,,
\ee
where $M_{\rm B}$, $J=a M_{\rm B}$, and $A_h$ are the total mass (the asymptotic value of the Bondi mass), total angular momentum and area of the event horizon, respectively. Following these three points: 1.~given that gravitational waves carry positive energy \cite{Bondi:1962px,Sachs:1962wk}, the Bondi mass can never be larger than the Arnowitt-Deser-Misner (ADM) mass $M$; 2.~the second law of black hole thermodynamics state that area of the event horizon cannot decrease \cite{Schon:1981vd,Bardeen:1973gs}; and 3.~any apparent horizon must be hidden inside the event horizon \cite{Hawking:1973uf}, and using Eq.~(\ref{kerr}), one can derive the following inequality
\be\label{penroseineq}
2{M}\geq\sqrt{\fft{A[\sigma]}{4\pi}}\,,
\ee
where $A[\sigma]$ is the minimal area enclosing the apparent horizon $\sigma$. This is the famous ``Penrose inequality'', and it was proposed by Penrose in the early seventies as a thought experiment for the ``cosmic censorship conjecture.''  (see, e.g.~\cite{Bray:2003ns, Mars:2009cj}). The Penrose inequality also has a rigid statement: saturation occurs only in the Schwarzschild spacetime \cite{Ben-Dov:2004lmn}.

Although the mathematical expression in \eqref{penroseineq} is very ``simple,'' its proof is a challenging problem in mathematical relativity and there is hitherto no general proof. However, there are two notable special cases where rigorous proofs have been given. The first deals with the general static geometries by assuming the weak energy condition (WEC) \cite{14,15}, and the second deals with dynamic spacetimes but with spherical symmetry \cite{Hayward:1994bu} by assuming the dominant energy condition (DEC). This inequality has also been generalized to asymptotically anti-de Sitter spacetimes (AdS) \cite{Itkin:2011ph,Husain:2017cmj,Engelhardt:2019btp,Xiao:2022obq}. Generally, either the DEC or WEC is assumed to be the requirement in the proof of static or dynamic spacetimes \cite{Mars:2009cj}. However, previous studies showed that the null energy condition (NEC) is sufficient to ensure Penrose inequality, at least for the spherically-symmetric and static black holes\cite{Engelhardt:2019btp,Xiao:2022obq,Yang:2022yye}.

The Penrose inequality involves the apparent horizon of the initial data set, and it does not require that the spacetime is or will finally settle into a stationary black hole. It expresses a relationship between the energy and the size of the black hole it occupies. If we focus on a stationary black hole, then the area of the horizon presents the Bekenstein-Hawking entropy of the system, which can be expressed as follows
\begin{equation}\label{defS}
	S=\fft14 A\,.
\end{equation}
The Penrose inequality then becomes entropy bound for a system of given total energy.

In thermodynamic systems, in addition to entropy, temperature $T$ is also an important quantity. Furthermore, in Einstein's gravity theory of gravity, mass, entropy, and temperature are all geometric quantities solely determined by the metric of the black hole. Thus, we need to investigate whether we could obtain any new bounds that involve all of these quantities. The fact that the product $T S$ has the same dimension of energy suggests that it modifies the mass bound in the inequality. Ref.~\cite{Yang:2022yye} made the first attempt to address this issue obtained following inequality for spherically-symmetric and static black holes by assuming only NEC:
\begin{equation}\label{EETineq}
	3{M}\geq\sqrt{\frac{S}{\pi}}+2TS\,.
\end{equation}
As discussed in Ref.~\cite{Yang:2022yye}, this inequality is independent of the Penrose inequality and saturation appears also only in the Schwarzschild black hole.

However, these two inequalities are unsatisfactory since both of them are too slack for extremal or near extremal black holes.  For example, the extremal (zero temperature) Reissner-Nordstr\"{o}m (RN) black hole has $M_{\rm ext}= \sqrt{S/\pi}$, which is twice the saturation value of the Penrose inequality and thrice the value of the inequality \eqref{EETineq}. An interesting question arises: Could we find a new lower bound for the total energy such that it is tighter than the inequalities \eqref{EETineq} and \eqref{penroseineq}? 
In Einstein's theory of gravity, the Einstein field equation relates the spacetime curvature to the energy-momentum tensor of minimally coupled matter which couples with gravity through the metric only. The essence of the Penrose inequality is about how the appropriate energy-momentum tensor constrains the spacetime geometry. Thus, we should be able to utilize all the geometric information to obtain a new bound that is more refined and stringent. Given that the Maxwell field fulfills all of the energy conditions, it is thus not entirely unreasonable to expect that the RN black hole will have the lowest mass for a given temperature and entropy. Furthermore, since the vacuum itself satisfies all of the energy conditions, we also expect that the new bound will be saturated by the vacuum solution, i.e., the Schwarzschild black hole.

We observe from the RN black hole that
\be\label{rnmts}
M=\frac{4 Q^2+r_+^2}{2 r_+}\,, \qquad TS=\frac{r_+^2-4Q^2}{4 r_+}\,,
\ee
where $r_+=\sqrt{S/\pi}$ is the radius of the event horizon. It is easy to check $M=r_+-2TS$. Based on our arguments, we propose following Penrose-like inequality for a static black hole
\be\label{CI}
{M} \ge\sqrt{\frac{S}{\pi}}-2TS\,.
\ee
(As we will discuss subsequently, the positivity of the right-hand side is ensured by the WEC.) For the Schwarzschild black hole, its Smarr relation $M=2TS$ implies that the above inequality is also saturated. For the extremal case, we have $T=0$ and the above inequality provides a tighter bound for the ADM mass compared with the Penrose inequality.

We shall examine a variety of explicit examples to support the inequality~\eqref{CI} in Section 2. We also provide an interesting counterexample. Then we formulate a conjecture of the energy conditions that guarantee the inequality \eqref{CI}. Subsequently, we provide general proof in Section 3 for the spherically-symmetric case. We determine that the required energy condition is the combination of the NEC and the trace energy condition (TEC), which contains the WEC. This is not a particularly strong requirement for such a tighter inequality. In doing so, we review four different ``quasi-local masses'' for the most-frequently used local energy conditions, i.e., NEC, WEC, the strong energy condition (SEC) and DEC, and illustrate how every one of them will provide different lower bound on ADM mass and also different proofs of the ``positive mass theorem.'' Then we construct a new quasi-local mass that gives us \eqref{CI} under ``NEC$+$TEC.'' In Section 4, we generalize our inequality to higher dimensions, or to include rotations or the cosmological constant. We conclude our paper in Section 5.

\section{Examples}\label{Ex}

In this section we study a variety of examples in four dimensions to illustrate the validity of our inequality (\ref{CI}). We also provide an interesting counterexample. We focus on the spherically-symmetric and static black holes, of which the metrics have the following general form
\be\label{metric0}
\mathrm{d}s^2 =-f (r)\mathrm{d}t^2 +\frac{\mathrm{d}r^2}{f(r)} + \rho(r)^2 \mathrm{d}\Omega^2\,,
\ee
Here $\mathrm{d}\Omega^2$ is the standard metric of the round unit 2-sphere. We denote $r_+$ to be the location of the event horizon where $f(r_+)=0$. The radius of the horizon is then $\rho(r_+)$. By defining
\be\label{xe}
X=M + 2 T S - \rho(r_+)\,,
\ee
we can verify whether we have $X\ge 0$ for a variety of black holes. Given that our inequalities involve only the geometric quantities, we shall present only the metrics of all of the examples.

\subsection{Born-Infeld}

The spherically-symmetric and asymptotically-flat solutions in the Einstein-Born-Infeld (EBI) theory were given in \cite{Breton:2002td,Li:2016nll}. The solution has $\rho=r$ and the blackening function $f$ is derived as follows
\be
f= 1 - \fft{2M}{r} + \ft16 b^2 \Big(r^2 - \sqrt{\fft{Q^2}{b^2} + r^4}\Big)
+ \fft{Q^2}{3r^2}\, {}_2F_1(\ft14,\ft12;\ft54;-\ft{Q^2}{b^2 r^4})\,.
\ee
Here $b>0$ is the Born-Infeld coupling constant, and $(M,Q)$ are the mass and electric charge respectively. It is straightforward to obtain
\be
T S=\frac{1}{8} r_+ \left(-b \sqrt{b^2 r_+^4+Q^2}+b^2 r_+^2+2\right).
\ee
Therefore we have
\be
\fft{6X}{b^2 r_+^3}=q^2 \, _2F_1\left(\frac{1}{4},\frac{1}{2};\frac{5}{4};-q^2\right)-2 \sqrt{q^2+1}+2\ge 0\,,
\ee
where $q\ge 0$ is the dimensionless parameter defined by $Q=q b r_+^2$. The inequality is saturated by $q=0$, which can be achieved by either $Q=0$, corresponding to the Schwarzschild black hole, or by $b\rightarrow \infty$, corresponding to the RN black hole.

\subsection{Quasitopological electromagnetism}

Given the theory in which the quasitopological electromagnetism is minimally coupled with gravity, one can find the exact solutions to the general spherically-symmetric and static dyonic black holes \cite{Liu:2019rib} that have $\rho=r$ and
\be
f=1 - \fft{2M}{r} + \fft{p^2}{r^2} +
\fft{q^2}{r^2} {}_2F_1\Big(\ft14,1;\ft54;-\ft{4\alpha p^2}{r^4}\Big),
\ee
where the three integration constants $(M, q, p)$ are associated with the mass, electric and magnetic charges. Moreover the black hole fulfills at
least the DEC once the coupling constant $\alpha$ is positive. This solution yields following expression
\be
T S=\frac{4 \alpha  p^4+r_+^4 \left(p^2+q^2\right)-4 \alpha  p^2 r_+^2-r_+^6}{4 r_+\left(4 \alpha  p^2+r_+^4\right)}.
\ee
Therefore, the following inequality can be obtained
\be
X=\frac{q^2 \left(\left(4 \tilde p^2+1\right) \, _2F_1\left(\frac{1}{4},1;\frac{5}{4};-4 \tilde p^2\right)-1\right)}{2 \left(4 \tilde p^2+1\right) r_+}\ge 0\,,
\ee
where $\tilde p = \alpha p/r_+^2>0$. The saturation happens when $q=0$, $p=0$ or $\alpha=0$, corresponding to the magnetic, electric or dyonic RN black holes respectively.

\subsection{STU supergravity model}

The $D = 4$, $N = 2$ STU  supergravity model coupled to three vector multiples admits four-charge
black hole solutions, expressed as follows (e.g.~\cite{Cvetic:1999xp})
\be
f=\fft{f_0}{\sqrt{H_1 H_2 H_3 H_4}}\,,\quad
f_0= 1 - \fft{\mu}{r}\,,\quad H_i = 1 + \fft{q_i}{r}\,,\quad
\rho=r (H_1 H_2 H_3 H_4)^{\fft14}\,,
\ee
where the solution is specified by five integration constants, including the electric charge parameters $q_i$ and the blackening parameter $\mu=r_+$. We derive the following equation
\be
M=\ft12\mu + \ft14 (q_1 + q_2 + q_3 + q_4)\,,\qquad T S = \ft1{4} r_+\,.
\ee
Thus we can easily verify
\be
X=\ft14 (\tilde q_1 + \tilde q_2 + \tilde q_3 + \tilde q_4 -
4 (\tilde q_1 \tilde q_2 \tilde q_3 \tilde q_4)^{\fft14}\ge 0\,,
\ee
where $\tilde q_i = q_i + r_+>0$.

\subsection{Einstein-scalar theory}

Here a scalar-hairy black hole in Einstein gravity is minimally coupled with a scalar with a specific scalar potential as discussed in \cite{Zhang:2014sta}. The metric is specified as follows
\be
f=1 + \alpha^2 \Big(r(q+r) \log \Big(\fft{r+q}{r}\Big) - \fft12
q(q+2r)\Big)\,,\qquad \rho = \sqrt{r(r+q)}\,.
\ee
In this case, we have
\be
M=\ft1{12} \alpha^2 q^3\,,\qquad T S = \ft14\alpha^2
r_+ \left(q+r_+\right) \left(\left(q+2 r_+\right) \log \left(\frac{q+r_+}{r_+}\right)-2 q\right)\,.
\ee
Thus Eq.~(\ref{xe}) leads to  ($x=r_+/q>0$)
\bea
\fft{X}{q}&=&-x-\sqrt{x} \sqrt{x+1}-\frac{2}{3 \left(-2 x+2 (x+1) x \log \left(\frac{1}{x}+1\right)-1\right)}-\frac{1}{2}\nn\\
&=&
\left\{
\begin{array}{ll}
	\fft16 -\sqrt{x}, &\qquad x\rightarrow 0\,; \\
	\fft{1}{40x}, &\qquad  x\rightarrow \infty\,.
\end{array}
\right.
\eea
Here $X$ is a monotonously-decreasing function from 1/6 to 0.

\subsection{Bardeen: a counterexample}

As our final example here, the Bardeen spacetime metric can be written as follows \cite{bardeen}
\be
f= 1 - \fft{2M r^2}{(r^2 + q^2)^{\fft32}}\,,\qquad \rho=r\,.
\ee
where $q$ is the magnetic charge. For fixed $q$, the extremal black hole has
\be
M_{\rm ext} = \ft34\sqrt3 q\,,\qquad r_+^{\rm ext} = \sqrt2 q\,.
\ee
Given $M_{\rm ext}< r_+^{\rm ext}$, we see this example violates the inequality (\ref{CI}). Since the Bardeen black hole satisfies WEC, but not the DEC or TEC, the WEC, which is sufficient for the Penrose inequality, is not enough to guarantee our inequality \eqref{CI}.

\subsection{Conjecture}
Although our examples are far from exhaustive, they represent different classes of matter fields. Based on these examples, we formulate the following conjecture:

\textit{For a four-dimensional asymptotically flat and static black hole in Einstein theory of gravity with the minimally coupled matter, if (1) the NEC and TEC are both satisfied and (2) the cross-section of the event horizon has $S^2$ topology, then the inequality \eqref{CI} is true}.

In the next section, we will provide a proof for the special spherically-symmetric case. Before we do that, we make a few comments.
\begin{itemize}

\item Our conjecture requires NEC+TEC. This combination implies the WEC, but not the DEC or SEC, which can however be consistently further imposed without upsetting the NEC+TEC.

\item The Penrose inequality turns out to be a corollary of our inequality~\eqref{CI}. To see this, we note that by assuming WEC, Refs.~\cite{Visser:1992qh,Yang:2020ark} have established following inequality
\begin{equation}\label{bdonts2}
	4TS\leq \sqrt{\fft{S}{\pi}}\,.
\end{equation}
This inequality not only ensures that the right-hand side of \eqref{CI} is positive but also helps us to deduce the Penrose inequality from \eqref{CI}, i.e., 
\be\label{CI2}
2M\ge 2\Big(\sqrt{\fft{S}{\pi}} - 2TS\Big)\geq \sqrt{\fft{S}{\pi}}=\sqrt{\fft{A}{4\pi}}\,.
\ee

\item Thirdly, given the temperature is always nonnegative for a static black hole, we derive the following inequality for extremal black holes,
\begin{equation}\label{zertot1}
	M_{\rm ext} \geq\sqrt{\fft{S}{4\pi}}=\sqrt{\fft{A}{4\pi}}\,.
\end{equation}
This gives us a tighter bound on the ADM mass for an extremal black hole, which cannot be improved anymore in the sense that it could be saturated by the extremal RN black hole.

\item Our inequality \eqref{CI} is tighter than the Penrose inequality and the recently proposed inequality \eqref{EETineq} is slaker than Penrose inequality. However, it is interesting to observe that the sum of these two inequalities gives precisely the Penrose inequality.

\end{itemize}

\section{Proof for spherically symmetric static black holes}

\subsection{Setup}

After examining the explicit examples, we now give a proof for a special class of black holes that are {\it static and spherically symmetric.} The most general such metrics takes following form
\be\label{metric}
\mathrm{d}s^2 =-f (r) e^{-\chi(r)} \mathrm{d}t^2 +\frac{\mathrm{d}r^2}{f(r)} +r^2 \mathrm{d}\Omega^2.
\ee
Asymptotic flatness requires that the functions $f$ and $\chi$ have the following falloff behaviors near infinity
\begin{equation}\label{asymfchi}
	f(r)=1-2M/r+\cdots\,,\qquad \lim_{r\rightarrow\infty}r\,\chi(r)=0\,,
\end{equation}
where $M$ is the ADM mass. Denoting $r_+$ as the radius of the event horizon, the temperature $T$ and entropy $S$ are
\be\label{ET}
T=\frac{f'(r_+)e^{-\chi(r_+)/2}}{4\pi}\,,\qquad S=\pi r_+^2\,.
\ee
For the matter energy momentum tensor
$
{T^\mu}_\nu=\text{diag}\{-\rho(r), p_{r}(r), p_{{T}}(r), p_{{T}} (r)\},
$ and the metric (\ref{metric}), the Einstein's equation leads to three ordinary differential equations
\begin{align}\label{Ein}
	f'&=\frac{1- 8 \pi r^2 \rho-f(r)} {r}\,,\qquad \chi'=\frac{-8\pi r (\rho +p_{r})}{f}\, ,\nonumber\\
	p'_r&=\frac{\rho-3p_r+4p_T}{2r} -\frac{(\rho+p_r)(1+8\pi p_r r^2)}{2 f r}\,.
\end{align}
In the next we will use these equations to prove the inequality (\ref{CI}). In fact, from the second equation, we observe that the NEC ensures that $\chi'\leq0$. Given that the asymptotic flatness requires $\chi(\infty)=0$, we conclude that the NEC forces $\chi\geq0$ outside of the horizon.

\subsection{Quasi-local masses and proof of inequality}

Our main tool is a new quasi-local mass. Before we present it, we provide a brief review of some quasi-local masses widely used in the proof of inequalities involving black hole mass. For the spherically-symmetric and static configurations, there exists four known quasi-local masses associated with the NEC, WEC, SEC, and DEC. These quasi-local masses are nondecreasing under the corresponding energy conditions and will generate different lower bounds for ADM mass, as well as different proofs on the ``positive mass theorem'' of black holes.

The first quasilocal  is the well-known ``Hawking-Geroch mass''~\cite{Hawking2,Geroch}, which plays crucial role in the proof of ``Penrose inequality''. For black holes of the metric \eqref{metric}, it has a rather simple form
\begin{align}\label{defm1}
	m_w(r)=\frac{r}2(1-f)\,.
\end{align}
Evaluating the Hawking-Geroch mass on the horizon and at infinity, one obtains the following equation
\begin{align}\label{defm1a}
	m_w(r_+)=\fft12 r_+\,,\qquad m_w(\infty)={M}\,.
\end{align}
Using the first equation in \eqref{Ein}, one finds
\begin{equation}\label{defm1b}
	m'_w=4 \pi r^2 \rho\,.
\end{equation}
We see that the WEC ensures the nondecreasing of Hawking-Geroch mass, from which the Penrose inequality arises naturally; hence the positive mass theorem for spherically-symmetric and static black holes is established.

Since here we consider the static black hole, the event horizon is a Killing horizon and there is a corresponding time-like Killing vector $\xi^\mu$ outside the black hole. The second quasi-local mass is the Komar mass, which in general is defined by the Komar integration over a closed 2-surface $\mathcal{S}$
\begin{equation}\label{defm2}
 m_s(\mathcal{S})=-\frac1{8\pi}\oint_{\mathcal{S}}(\mathrm{d}\xi)_{\mu\nu}\mathrm{d}S^{\mu\nu}\,.
\end{equation}
By specifying  the spherically-symmetric case under the coordinate gauge~\eqref{metric}, with an equal-$r$ sphere $\mathcal{S}$ , we have
\begin{equation}\label{defm2a}
	m_s(r)=\frac12r^2e^{\chi/2}(fe^{-\chi})'\,.
\end{equation}
Using the Komar mass we can easily obtain a positive mass theorem for general static and asymptotically flat black holes, not only for the special spherically-symmetric case. Denote $\Sigma$ to be a 3-dimensional hypersurface outside the event horizon with a boundary $H$ at the Killing horizon and the boundary $\mathcal{S}_\infty$ at the spatial infinity. We chose $\Sigma$ to be orthogonal to the Killing vector $\xi^\mu$. We have
\begin{equation}\label{defm2b}
	m_s(\mathcal{S}_\infty)=M\,,\qquad m_s(H)=\fft12 TS\,,
\end{equation}
and
\begin{equation}\label{defm2d} m_s(\mathcal{S}_\infty)-m_s(H)=-\frac1{8\pi}\oint_{\partial\Sigma}
(\mathrm{d}\xi)_{\mu\nu}\mathrm{d}S^{\mu\nu}=2\int_{\Sigma}(T_{\mu\nu}-g_{\mu\nu}T/2)\xi^\mu n^\mu\mathrm{d} V\,.
\end{equation}
Here $n^\mu$ is the unit normal vector of $\Sigma$ and $\xi^\mu\parallel n^\mu$ with $\xi^\mu n_\mu<0$. From this we can easily see that the SEC can ensure the nonnegativity of $m_s(\mathcal{S}_\infty)-m_s(H)$. Thus, we obtain the following inequality under the assumption of the SEC
\begin{equation}\label{inequalitydec}
	M\geq2TS\,.
\end{equation}
This inequality provides proof for the ``positive mass theorem'' under the assumption of the SEC. (It is worth commenting here that our inequality \eqref{CI}, together with \eqref{bdonts2}, implies that
\eqref{inequalitydec} can also be satisfied with NEC+TEC, which is independent of the SEC.)

The third quasi-local mass is very similar to Hawking-Georch mass, and is defined as follows
\begin{equation}\label{modifhawingm1}
	m_{d}(r) =\frac{r}2(1 - e^{-\chi/2} f)\,.
\end{equation}
This and Hawking-Georch mass have the same values on the event horizon and at infinity. However, their derivatives with respective to $r$ are different. Using both the first and second equations in \eqref{Ein}, we obtain
\begin{equation}\label{modfihawkingm2}
	m_d' =\frac12(1 - e^{-\chi/2}) + 2\pi r^2 e^{-\chi/2}(\rho-p_r)\,.
\end{equation}
We see that the DEC can ensure $m_d'\geq0$ and therefore it provides a different proof on Penrose inequality, as well as the ``positive mass theorem'' for spherically symmetric black holes

The fourth quasi-local mass was recently proposed in Refs.~\cite{Xiao:2022obq,Yang:2022yye}. Although it currently applies only to the spherically-symmetric and static black holes, the result is surprising since it establishes the Penrose inequality and the ``positive mass theorem'' by assuming NEC only. Specifically, it can be expressed as follows
\begin{equation}\label{defmnec}
	m_n(r)=\frac{r^4e^{\chi/2}}6\left(\frac{fe^{-\chi}}{r^2}\right)'+\frac{r}3\,.
\end{equation}
Evaluating this inequality on the horizon and at infinity produces the following results
\begin{equation}\label{defmnec2}
	m_n(r_+)=\frac{2TS}{3}+\frac{r_+}3\,,\qquad m_n(\infty)=M\,.
\end{equation}
Using all of the three terms in \eqref{Ein}, one finds
\begin{equation}\label{defmnec3}
	m_n'=\frac{8\pi}3 e^{-\chi/2}r^2(\rho+p_T)+\frac13[1-e^{-\chi(r)/2}]\,.
\end{equation}
It follows that NEC can ensure $m_n$ is nondecreasing; thus, we obtain the inequality \eqref{EETineq}. Interestingly, using this quasi-local mass one can further prove the Penrose inequality \cite{Xiao:2022obq,Yang:2022yye}.

In this study, we will propose a new quasi-local mass, which is compatible with TEC and will be used to prove our new inequality~\eqref{CI}. To do so, we define a quasi-local mass that is the most general linear combination of $m_n, m_w, m_s$, and $m_d$ such that
\be\label{atih}
m(\infty) = {M}\,,\qquad m(r_+) = r_+ - 2 TS\,.
\ee
We determine that the combination has two free parameters $(\alpha, \beta)$:
\be
m(r) = 2m_w(r) - m_s(r) + \alpha (3m_n(r)-m_s(r) - 2m_w(r)) + 2\beta (m_d - m_w)\,.\label{result}
\ee
Its derivative yields
\be\label{pmf}
m'(r) = \gamma (1-e^{-\fft12\chi}) + (1-\gamma) \pi r^2 \rho (1-e^{-\fft12\chi})
+ 4\pi r^2 e^{-\fft12 \chi} \Big((1-\gamma)\rho - (1+\gamma) p_r - 2p_T\Big)\,.
\ee
Here $\gamma=\alpha+\beta$. For $m'(r) \ge 0$, the first and second terms of the above equation suggest that $\gamma\in [0,1]$ and $\rho \ge 0$ which is WEC. The NEC implies $\chi'\le 0$. Thus, from the asymptotic flatness we have $\chi\ge 0$ and $0<e^{-\chi(r)/2}\le 1$. If we are only allowed to impose any of the five energy conditions, but nothing more, then it appears that we have to set $\gamma=\alpha+\beta=0$. Therefore, from Eq.~(\ref{pmf}) we have
\be
m'_{\tr}(r)\ge 4 \pi r^2 (\rho-p_r-2p_T) = 4 \pi r^2 (-T),
\ee
in which $T$ denotes the trace of the energy momentum tensor. (There should be no confusion with the temperature.) By imposing TEC ($-T\ge0$) to the above relation we have $m'_{\tr}(r) \ge 0$. Therefore we obtain a one-parameter family of quasi-local masses \eqref{result} with $\alpha+\beta=0$ that is nondecreasing under NEC$+$TEC.  Finally, from Eq.~(\ref{atih}) we conclude that the inequality (\ref{CI}) is true when the NEC$+$TEC is satisfied. Note that the WEC will also be satisfied under requirement of NEC+TEC.

Finally, we would like to discuss whether our new quasi-local mass associated with the TEC could also apply to the general static black holes. If we set $\alpha=0=\beta$, then the quasi-local mass is simply a linear combination of the Hawking-Geroch and Komar masses, both of which can be extended to cover general static configurations without spherical symmetry. We denote $t$ as the time coordinate of the static spacetime and consider a 3-dimensional equal-$t$ spacelike hypersurface $\Sigma$ outside the event horizon. Taking $\mathcal{S}\subset\Sigma$ as  an arbitrarily closed 2-dimensional surface, the Komar mass is still given by Eq.~\eqref{defm2} and the Hawking-Geroch mass is now defined as follows
\begin{equation}\label{generalHGm}
  m_w(\mathcal{S})=\sqrt{\frac{A(\mathcal{S})}{16\pi}}
  \left(1-\frac1{16\pi}\int_{\mathcal{S}}k^2\mathrm{d} S\right),
\end{equation}
where $A(\mathcal{S})$ and $k$ are the area and extrinsic curvature (embedded in $\Sigma$) of surface $\mathcal{S}$, respectively. Thus, our new quasi-local mass for general static spacetimes can be expressed as follows
\begin{equation}\label{generalmt}
  m(\mathcal{S})=\sqrt{\frac{A(\mathcal{S})}{4\pi}}\left(1-\frac1{16\pi}\int_{\mathcal{S}}k^2\mathrm{d} S\right)+\frac1{8\pi}\oint_{\mathcal{S}}(\mathrm{d}\xi)_{\mu\nu}\mathrm{d}S^{\mu\nu}.
\end{equation}
On the event horizon, with $k=0$, we have
\begin{equation}\label{generalmth}
  m(H)=\sqrt{\frac{S}{\pi}}-2TS\,.
\end{equation}
At infinity we have $m(\infty)=M$. It is of great interest to study whether the NEC$+$TEC that is sufficient for the spherically-symmetric cases can also ensure $m(\infty)\geq m(H)$ for the more general static configurations.

\section{Generalizations}

\subsection{To higher dimensions}

Our conjecture and its proof are narrow, focusing only on the spherically-symmetric and static black holes in four dimensions.  Nevertheless, its success prompts us to consider its possible generalizations. The simplest is perhaps to higher dimensions. For static black holes, we expect
\be
M + \fft{D-2}{D-3} T S \ge \fft{D-2}{\pi} \Big(\fft{\Omega_{D-2}}{2^D}\Big)^{\fft1{D-2}}
S^{\fft{D-3}{D-2}}\,,
\ee
where $\Omega_k=2\pi^{\fft{k+1}2}/(\fft12 (k-1))!$ is the volume of the round unit $k$-sphere. This inequality is again saturated by Schwarzschild-Tangherlini and RN-Tangherlini black holes.  For spherically-symmetric black holes, the proof is straightforward.

\subsection{To rotations}

We now consider rotating black holes and for simplicity, we consider only four dimensions. We can easily check that the Kerr black hole does not satisfy the inequality \eqref{CI} as follows
\be
(M + 2 T S)^2 - \fft{S}{\pi}=-a^2<0\,.\label{Kerrvio}
\ee
This perhaps is unsurprising since as we explained in the ``Introduction'', a tight inequality should utilize all of the black hole geometric information, which certainly includes the angular momentum $J$ and its conjugate velocity $\Omega$, when it is rotating. Based on the dimensions of these quantities, we can have two simple approaches to modify Eq. \eqref{CI}. The first is to take advantage of the fact that $\Omega J$ has the same dimension as the mass, and it is natural to propose that
\be
M + 2 T S+ \eta\, \Omega J \ge \sqrt{\fft{S}{\pi}}\,,\label{rotineq}
\ee
for a certain order-one purely numerical constant $\eta$. Since we expect that $\Omega J$ is positive, a higher positive $\eta$ value makes the inequality more trivial.  Therefore, the key is to find the minimum value of $\eta$. The examples that we will discuss suggest that $\eta \ge 1$.

In the second approach, we note that angular momentum $J$ has the same dimension as entropy, we can modify \eqref{CI} as follows
\be
M + 2 T S \ge \sqrt{\fft{S}{\pi} - |J|}\,.\label{rotineq2}
\ee
In other words, we made the right-hand side of \eqref{CI} a slightly smaller. The precise ``$-1$'' factor in front of $|J|$ is determined by requiring that the inequality is saturated by the extremal Kerr black hole. An inequality established in Ref.~\cite{Dain:2011pi}, i.e., 
\be
\fft{S}{\pi} \ge 2 |J|\,,
\ee
ensures that the right-hand side of Eq. \eqref{rotineq2} is always a real and positive number. Thus, the full conjecture is expressed as follows
\be
M + 2 T S \ge \sqrt{\fft{S}{\pi} - |J|}\,\ge \sqrt{|J|}\,.\label{rotineq3}
\ee

\subsubsection{Kerr-Newmann black hole}

We first examine the Kerr-Newmann black hole and its angular velocity, temperature and entropy are \cite{4}
\be
\Omega = \fft{a}{r_+^2 + a^2}\,,\qquad T=\fft{r_+-M}{2\pi(r_+^2 + a^2)}\,,\qquad
S=\pi (r_+^2 + a^2)\,,
\ee
where the parameters $M, Q$, and $a = J/M$ are the mass, charge and the angular momentum (per mass) respectively. The mass $M$ is related to the horizon location by
$r_+^2 + a^2 + Q^2 -2 M r_+=0.$ Thus, it is easy to verify that for $\eta=0$, we have \eqref{Kerrvio}, independent of the charge. However, we find that Eq. \eqref{rotineq} can be satisfied for $\eta\ge 1$. Specifically, we derive the following expression
\be
(M + 2 T S + \Omega J)^2 - \fft{S}{\pi}=\frac{2 r_+^2 \left(a^6+3 a^4 Q^2\right)+r_+^4 \left(a^4+4 a^2 Q^2\right)+a^4 \left(a^2+Q^2\right)^2}{4 r_+^2 \left(a^2+r_+^2\right)^2}>0\,.
\ee
To test the sequence of inequalities in Eq. \eqref{rotineq3}, we find
\bea
&&(M + 2 T S)^2 - \Big(\fft{S}{\pi}-|J|\Big) = \frac{|a| \left(\left(r_+-a\right){}^2+Q^2\right)}{2 r_+}\ge 0\,,\nn\\
&& \fft{S}{\pi} - 2 |J|= 4 \pi  |a| \left(a^2+r_+^2\right) T+\left(r_+-a\right)^2\ge 0\,.
\eea

\subsubsection{Kerr-Sen black hole}

The Kerr-Sen black hole was generated from the Kerr black hole to carry an electric charge in the string theory \cite{Sen:1992ua}.  The solution contains three parameters, the mass $M$, angular momentum $J$, and the electric charge $Q$. Temperature, entropy and angular velocity can be expressed as follows
\bea\label{kerrsents}
T&=&\frac{\sqrt{\left(2 M^2-Q^2\right)^2-4 J^2}}{4 \pi  M \left(\sqrt{\left(2 M^2-Q^2\right)^2-4 J^2}+2 M^2-Q^2\right)}\,,\nn\\
S &=& 2 \pi  M \left(\sqrt{\left(M-\frac{Q^2}{2 M}\right)^2-\frac{J^2}{M^2}}-\frac{Q^2}{2 M}+M\right),\nn\\
\Omega &=&\frac{J}{2 M^2 \left(\sqrt{\left(M-\frac{Q^2}{2 M}\right)^2-\frac{J^2}{M^2}}-\frac{Q^2}{2 M}+M\right)}\,.
\eea
We determine that together with the electric potential $\Phi=Q/(2M)$, the first law $dM=TdS + \Omega dJ + \Phi dQ$ is satisfied. Notably, the theory considered in Ref. \cite{Sen:1992ua} was in the string flame, and the dilaton was not minimally coupled. However, one can perform a conformal transformation such that all the matter fields become minimally coupled. Under conformal transformation, all the thermodynamic variables and the first law remain the same. Although the expressions in \eqref{kerrsents} are complicated, we can prove the inequalities analytically. The parameters must satisfy the following: 
\be
M\ge M_{\rm ext} =\sqrt{|J| + \ft12 Q^2}\,.
\ee
For $\eta\ge 1$, we find that the inequality \eqref{rotineq} is satisfied. Specifically, we derive the following equation
\be
(M + 2 T S + \Omega J)^2 - \fft{S}{\pi} = \fft{\Big(2M^2 + Q^2 - 2 \sqrt{(M^2- \ft12 Q^2)^2 - J^2}\Big)^2}{16M^2} \ge 0\,.
\ee
We also examine the sequence of the inequalities in \eqref{rotineq3} and determine that
\bea
&&(M + 2 T S)^2 - \Big(\fft{S}{\pi}-|J|\Big) = \fft{1}{4M^2}\Bigg( Q^4 + 2 |J|
\Big(Q^2 + 2(M^2-\ft12 Q^2- |J|)\Big) \Bigg)\ge 0\,,\nn\\
&& \fft{S}{\pi} - 2 |J|= 2(M^2- \ft12 Q^2 - |J|) +
2\sqrt{(M^2 - \ft12 Q^2)^2 - J^2}\ge 0\,.
\eea
Notably, that the second inequality is saturated by the extremal Kerr-Sen solution.

There are more complicated and involved rotating black holes, such as the four-charge
Cveti\v c-Youm black hole \cite{Cvetic:1996kv,Chong:2004na}. We shall investigate the inequalities of these black holes in future studies.

\subsection{To include the cosmological constant}

For simplicity, we consider the cosmological constant $\Lambda$ minimally coupled with all fields in theory. We also assume that the black hole is static and spherically symmetric, and the leading metric falloffs at the asymptotic infinity are associated with the mass term. For the negative cosmological constant, the black hole is asymptotic to AdS. In this case, for given $M$, the horizon becomes smaller and the temperature becomes higher than the case without the cosmological constant. Thus the inequality becomes easier to satisfy, without saturation. This suggests that we have not taken into account all of the geometric information associated with the AdS black holes. Previous studies proposed that the cosmological constant can be viewed as the thermodynamic pressure $P_{\rm th}$ and there is a conjugate thermodynamic volume $V_{\rm th}$
\cite{Kastor:2009wy,Cvetic:2010jb}.  Their product $P_{\rm th} V_{\rm th}$ is an energy-like quantity.  Therefor, we can propose that
\be
M + 2 T S+ \xi\, P_{\rm th} V_{\rm th} \ge \sqrt{\fft{S}{\pi}}\,,\label{cosmoineq0}
\ee
for some order-one purely numerical constant $\xi$. In the simple case of a minimally coupled cosmological constant $\Lambda$, we have
\be
P_{\rm th}= - \fft{\Lambda}{8\pi}\,,\qquad V_{\rm th} = \fft43 \pi r_+^3\,,
\ee
where $r_+$ is the radius of the black hole horizon. We find that the proposed inequality \eqref{cosmoineq0} can be saturated with $\xi = -4$ by both the Schwarzschild-(A)dS and RN-(A)dS black holes, yielding
\be
M + 2 T S -4 \, P_{\rm th} V_{\rm th} \ge \sqrt{\fft{S}{\pi}}\,.\label{cosmoineq1}
\ee
It is straightforward to establish that the asymptotic (A)dS black holes in EBI and Einstein quasitopological electromagnetism discussed previously also satisfy the inequality. For a negative cosmological constant, the $P_{\rm th} V_{\rm th}$ term tightens the inequality. Meanwhile. for a positive cosmological constant, $P_{\rm th} V_{\rm th}$ term  increases the gap for the easier satisfaction of the inequality.

When the cosmological constant couples with scalars non-minimally, e.g., gauge supergravities, the situation becomes more complicated, because the definition $V_{\rm th}$ can be ambiguous \cite{Feng:2017wvc}. It is a subject worth further study.

\section{Conclusion}

Motivated by the observation that the Penrose inequality for extremal black holes have a large gap from saturation, we proposed a tighter bound (\ref{CI}) for static black holes. Our inequality contains all of the geometric information associated with energy-like quantities, including the mass and the product of temperature and entropy. Therefore, it is a more refined inequality.

Firstly, we tested our inequality with a variety of concrete examples of the known black holes in the literature, and the results are summarized in the Table~\ref{tab:table1}. We then provided a proof for general spherically-symmetric and static black holes. We combined the known four ``quasi-local" masses and obtained a new one, which enabled us to show that the inequality (\ref{CI}) was true when both NEC and TEC were satisfied. We also showed that our inequality~\eqref{CI} was always stronger than the Penrose inequality in the sense that the latter was a corollary of \eqref{CI}. Here we emphasize that the requirement ``NEC+TEC'' is a sufficient condition. As we have summarized in Table~\ref{tab:table1} that the even though TEC can be violated in the Einstein-scalar model, the inequality \eqref{CI} remains nevertheless true.

\begingroup
\begin{table}
	\begin{center}
		\begin{tabular}{c | c | c| c | c | c|c }
			\textbf{BH/EC} & \textbf{NEC} & \textbf{WEC} & \textbf{DEC} & \textbf{SEC} & \textbf{NEC+TEC}&inequality~\eqref{CI} \\
			\hline
			\multirow{1}{*}{RN} & True & True & True & True & True&True \\
			\hline
			\multirow{1}{*}{Born-Infeld} & True & True & True & True &True&True
			\\
			\hline
			\multirow{1}{*}{QTE ($\alpha \ge 0)$} & True & True & True & True & True& True
			\\
			\hline
			\multirow{1}{*}{STU} & True & True & True & True & True& True
			\\
			\hline
			\multirow{1}{*}{Pure scalar} & True & False  &  False & False & False& True
			\\
			\hline
			\multirow{1}{*}{Bardeen} & True & True & False & False & False& False
		\end{tabular}
\caption{Summary of the concrete examples}
		\label{tab:table1}
	\end{center}
\end{table}
\endgroup

It is perhaps physically more significant to rewrite the inequalities \eqref{EETineq} and \eqref{CI} as upper and lower bounds on the temperature for a nonrotating system of energy $M$ and entropy $S$, namely
\be
\sqrt{\fft{1}{\pi\, S}} - \fft{M}{S}\, \le 2T \le\, \fft{3M}{S} - \sqrt{\fft{1}{\pi S}}\,.\nn
\ee
As in the case of the Penrose inequality that provides a universal entropy upper bound for given energy, we expect that the above inequality will also provide a universal upper and lower bounds for the temperature of a nonrotating system.

In addition, we consider generalizations in various directions. The generalization of \eqref{CI} to higher dimensions for spherically-symmetric and static black holes is straightforward. A generalization to include rotation is more subtle. In this case, we need augment the inequality with additional information such as the product of angular momentum and its conjugate angular velocity. We find that the inequality \eqref{rotineq} is satisfied by the Kerr-Newmann and Kerr-Sen black holes provided that $\eta\ge 1$. Alternatively, since the dimensions of $J$ and $S$ are the same, it is also natural to propose the inequality \eqref{rotineq3}. We find that they are both satisfied by the Kerr-Newmann and Kerr-Sen solutions. For the second proposal, the bound is tight in that it is saturated by the extremal Kerr black hole. For asymptotic (A)dS spacetimes, we proposed \eqref{cosmoineq1} that is saturated by both the Schwarzschild-(A)dS or RN-(A)dS black holes. However, whether this inequality can stand against more examples or rigorous proof remains to be investigated.

One further issue needs to be addressed. It is well known that the Penrose inequality is saturated only by the Schwarzschild black hole, which becomes special from the Penrose inequality point of view. Our inequality is saturated by the RN black hole. Thus it is intriguing to ask whether this is the only nonvacauum solution that saturates inequality. To narrow down the statement further, is the extremal RN black hole is the only extremal one that saturates the bound expressed in Eq. \eqref{zertot1}? It will be interesting to investigate.

\section*{Acknowledgement}

This work is supported in part by the National Natural Science Foundation of China (NSFC) grants No.12005155, No.~11875200 and No.~11935009.

\end{document}